\documentclass[journal=NanoLetters,manuscript=article]{achemso}
\setkeys{acs}{keywords = true}

\usepackage[version=3]{mhchem} 
\usepackage{graphicx}
\usepackage{subfigure}
\usepackage{epsfig}
\usepackage{xcolor}
\usepackage{dcolumn}
\usepackage{bm}
\usepackage{ulem}
\usepackage{color}
\usepackage{amsmath}
\usepackage{amsfonts}
\usepackage{amssymb}
\usepackage{booktabs}
\usepackage{listings}
\usepackage{lmodern}
\usepackage{microtype}
\usepackage{natmove}
\usepackage{natbib}

\usepackage{epstopdf}

\def\be{\begin{equation}}       \def\ee{\end{equation}}
\def\bea{\begin{eqnarray}}      \def\eea{\end{eqnarray}}
\def\ba{\begin{array} }
\def\ea{\end{array} }
\def\bnum{\begin{enumerate} }
\def\enum{\end{enumerate}}

\def\=>{\Rightarrow}
\def\>{\rightarrow}

\def\eye2{Fathbb{I}}

\renewcommand{\>}{\rangle}

\author{Eryin Wang}
\altaffiliation{These authors contribute equally to this work.}
\affiliation{State Key Laboratory of Low Dimensional Quantum Physics and Department of Physics, Tsinghua University, Beijing 100084, P.R. China}

\author{Peizhe Tang}
\altaffiliation{These authors contribute equally to this work.}
\affiliation{State Key Laboratory of Low Dimensional Quantum Physics and Department of Physics, Tsinghua University, Beijing 100084, P.R. China}

\author{Guoliang Wan}
\affiliation{State Key Laboratory of Low Dimensional Quantum Physics and Department of Physics, Tsinghua University, Beijing 100084, P.R. China}

\author{Alexei V. Fedorov}
\affiliation{Advanced Light Source, Lawrence Berkeley National Laboratory, Berkeley, CA 94720, USA}

\author{Ireneusz Miotkowski}
\affiliation{Department of Physics and Astronomy, Purdue University, West Lafayette IN 47907, USA}

\author{Yong P. Chen}
\affiliation{Department of Physics and Astronomy, Purdue University, West Lafayette IN 47907, USA}
\alsoaffiliation{Birck Nanotechnology Center, Purdue University, West Lafayette IN 47907, USA}
\alsoaffiliation{School of Electrical and Computer Engineering, Purdue University, West Lafayette IN 47907, USA}

\author{Wenhui Duan}
\affiliation{State Key Laboratory of Low Dimensional Quantum Physics and Department of Physics, Tsinghua University, Beijing 100084, P.R. China}
\alsoaffiliation{Institute for Advanced Study, Tsinghua University, Beijing 100084, P.R. China}
\alsoaffiliation{Collaborative Innovation Center of Quantum Matter, Beijing 100084, P.R. China}
\email{dwh@phys.tsinghua.edu.cn}
\phone{+86 010 62772780}

\author{Shuyun Zhou}
\affiliation{State Key Laboratory of Low Dimensional Quantum Physics and Department of Physics, Tsinghua University, Beijing 100084, P.R. China}
\alsoaffiliation{Collaborative Innovation Center of Quantum Matter, Beijing 100084, P.R. China}
\email{syzhou@mail.tsinghua.edu.cn}
\phone{+86 010 62797928}

\title[An \textsf{achemso} demo]
  {Robust gapless surface state and Rashba-splitting bands  upon surface deposition of magnetic Cr on Bi$_2$Se$_3$ \footnote{The authors declare no competing financial interest.}}


\begin{document}

\begin{abstract}
  The interaction between magnetic impurities and the gapless surface state is of critical importance for realizing novel quantum phenomena and new functionalities in topological insulators.  By combining angle-resolved photoemission spectroscopic experiments with density functional theory calculations, we show that surface deposition of Cr atoms on Bi$_2$Se$_3$ does not lead to gap opening of the surface state at the Dirac point, indicating the absence of long-range out-of-plane ferromagnetism down to our measurement temperature of 15 K. This is in sharp contrast to bulk Cr doping, and the origin is attributed to different Cr occupation sites. These results highlight the importance of nanoscale configuration of doped magnetic impurities in determining the electronic and magnetic properties of topological insulators.
\end{abstract}




{Three dimensional (3D) topological insulator (TI), a novel class of materials  \cite{HasanRMP, SCZhangRMP}, is a topologically nontrivial bulk insulator with conducting surface state (SS). Protected by the time-reversal symmetry (TRS), the SS is spin-polarized, gapless and robust against nonmagnetic impurities \cite{AliScattering, XueScattering, KLWangNano1, DPYuNano}.  Meanwhile, magnetic impurities may induce out-of-plane ferromagnetism and break the TRS in TI, resulting in a  band gap at the Dirac point of the SS \cite{YLCdoping}. Such TRS breaking is essential for the realization of various novel quantum phenomena which have great application potential in future nano-electronic and spintronic devices, such as quantum anomalous Hall effect (QAHE) \cite{SCZhangQAH,XueQAH,KLWangPRL,TokuraNPhys}, topological magnetoelectric effect (TME) \cite{XLQFieldTl} and birefringent spin lens \cite{DuanLens}. Thus, it is critical to understand the interaction between 3D TIs and magnetic impurities both for fundamental physics and nano-device applications.

The interaction between 3D TI and doped magnetic impurities can modify the electronic and magnetic properties of 3D TI, and the results may strongly depend on the types of magnetic atoms, occupation sites of the magnetic impurities \cite{YaoYG} and experimental conditions under which they are introduced. For example, although doping Fe and Cr into the bulk Bi$_2$Se$_3$ crystal or thin film during the growth process leads to a gap opening at the Dirac point suggesting TRS breaking \cite{YLCdoping, XuePRL2014,YCuiNano1,YCuiNano2}, the robustness of the gapless SS against Fe deposition on the surface has been debated  \cite{Hasandoping,BiSe-Fe-doping-2} and surface deposition of Cr still remains to be investigated. Here we present a combined experimental and theoretical study of the electronic structures of Bi$_2$Se$_3$ upon surface deposition of magnetic Cr atoms on freshly cleaved Bi$_2$Se$_3$ single crystals. Angle-resolved photoemission spectroscopy (ARPES) data show that the SS is robust upon surface deposition of Cr, and the absence of gap opening at the Dirac point suggests that there is no long-range out-of-plane ferromagnetism down to the lowest measurement temperature of 15 K. In addition, Cr atoms dope electrons to the bulk states of Bi$_2$Se$_3$ forming a two-dimensional electron gas  with large Rashba-splitting on the surface. These observations are in striking contrast to the bulk doping of Cr during thin film growth \cite{XuePRL2014}, suggesting that Cr atoms deposited on the surface interact very differently with the TI sample.  Density functional theory (DFT) calculations show that the difference in electronic and magnetic properties originates from the different Cr configurations: hollow site for surface deposition versus substitutional site for bulk doping}. Due to the coupling between Cr 3d orbitals around the Fermi level, the magnetic easy axis of deposited Cr atoms is likely along the in-plane direction, which is in agreement with the robustness of the SS.  These results highlight the importance of nanoscale configuration of magnetic impurities in determining the electronic and magnetic properties of topological insulators, and such understanding is critical for manipulating the electronic properties of TI and future electronic applications.

\begin{figure}
\includegraphics[width=14 cm] {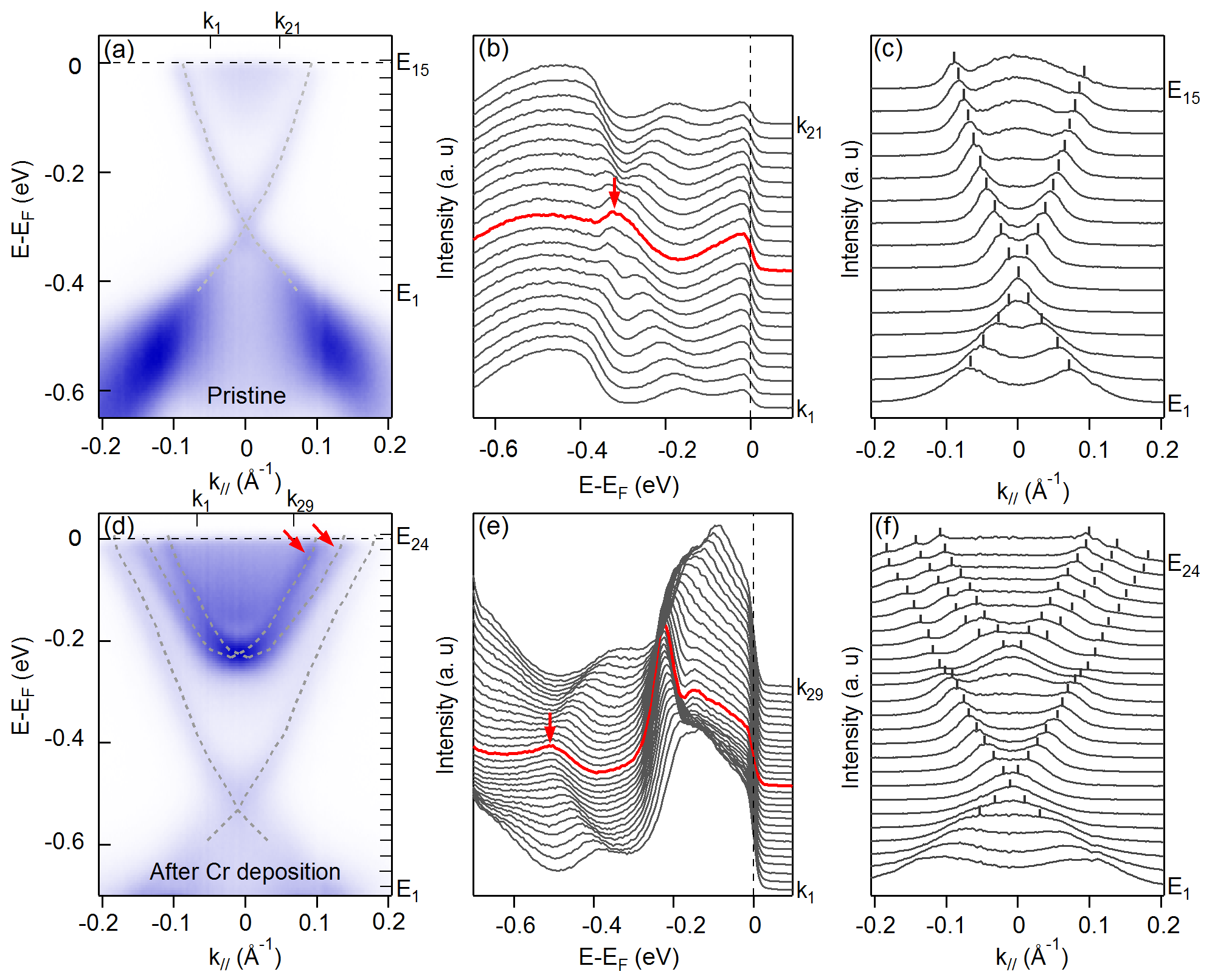}
\caption{(Color online) (a) Dispersion image along M-$\Gamma$-M direction. Dotted lines show dispersions extracted from MDCs in (c). (b) EDCs at momentum positions between k$_1$ and k$_{21}$ marked by ticks in (a). The EDC at the $\Gamma$ point is highlighted by red color. (c) MDCs at energy positions marked by ticks between E$_1$ and E$_{21}$ in (a). Ticks are guides for peak positions of the SS. (d) Dispersion image along M-$\Gamma$-M direction after surface deposition of Cr for 46 minutes, $\sim 9\%$ ML Cr on Bi$_2$Se$_3$. Dotted lines show dispersions extracted from MDCs in (f). (e) EDCs at momentum positions between k$_1$ and k$_{29}$ marked by ticks in (d). The EDC at $\Gamma$ point is highlighted by red color. (f) MDCs at energy positions marked by ticks between E$_1$ and E$_{24}$ in (d). Ticks are guides for peak positions of the SS and Rashba splitting bands.}
\label{fig:ARPES}
\end{figure}

Figure \ref{fig:ARPES} shows a comparison of ARPES data measured at 15 K with 35 eV photon energy on freshly cleaved Bi$_2$Se$_3$(111) surface (a) and after deposition of Cr atoms on the surface for 46 minutes, with coverage $\sim 9\%$ monoatomic layer (ML) Cr (d).  ARPES data on pristine  Bi$_2$Se$_3$  [Fig. \ref{fig:ARPES}(a)] shows conical dispersion from the SS with Dirac point energy at -0.3 eV, a small parabolic bulk conduction band near E$_F$. After deposition of Cr for 46 minutes, the SS persists and is electron doped, with the Dirac point energy shifted down to -0.51 eV. Energy distribution curves (EDCs) and momentum distribution curves (MDCs) analysis shows that there is no gap opening at the Dirac point of the SS down to 15 K [Fig.~\ref{fig:ARPES}(e, f)]. This is in striking difference from bulk doping of Cr into Bi$_2$Se$_3$ films during MBE growth \cite{XuePRL2014, YYWangCrdoping}, where a significant gap at the Dirac point was reported at a much higher temperature. Despite the magnetic moment of Cr atoms \cite{CrMagMom1, CrMagMom2, CrMagMom3} , our results suggest the surface state is very robust upon Cr surface deposition.

\begin{figure}
\includegraphics[width=16.8 cm] {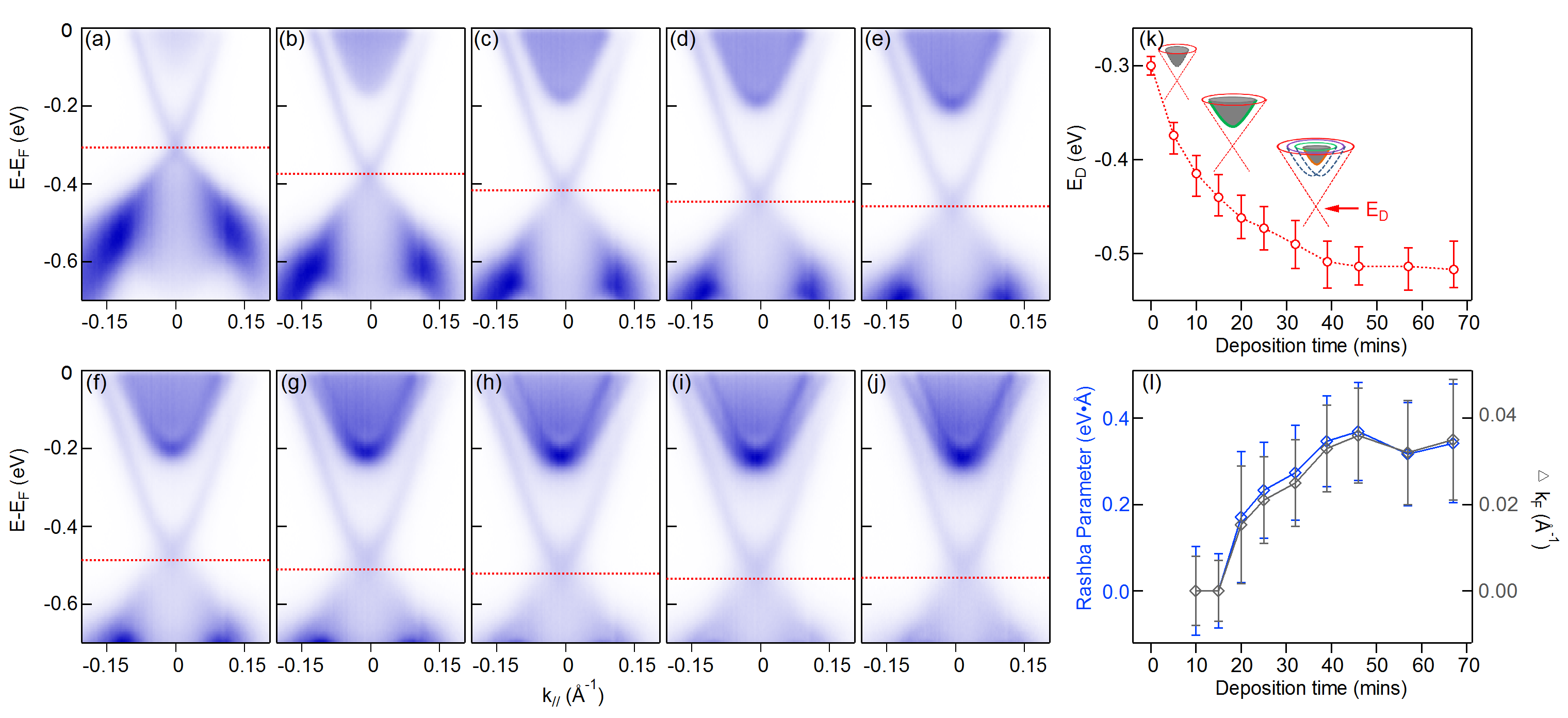}
\caption{(Color online) (a-j) Dispersions along M-$\Gamma$-M direction on pristine Bi$_2$Se$_3$ (a) and with different Cr deposition time 5, 10, 15, 20, 25, 32, 39, 46 and 57 minutes (b-j).  Red dotted lines indicate the energy positions of the Dirac point. (k) Evolution of the Dirac point energy as a function of Cr deposition time.  The inset shows a schematic illustration of the band structure evolution. (l) Rashba splitting $\Delta k_F$ (black symbols, right axis) and calculated Rashba coupling parameter $\alpha$ (blue symbols, left axis) as a function of Cr deposition time.}
\label{fig:doping}
\end{figure}

While doping electrons to the SS with preserved TRS, Cr deposition also induces additional bands (pointed to by arrows in Fig.~1(d)), which are attributed to two dimensional electron gas (2DEG) formed on the sample surface. Figures~\ref{fig:doping}(a-j) show the evolution of the band dispersion with increasing Cr deposition time. Initial deposition induces electron doping to the sample and shifts the SS and bulk bands to higher binding energy.  A parabolic band of 2DEG forms at the bottom of the bulk conduction band after depositing Cr for 10 minutes [Fig.~\ref{fig:doping}(c)]. With increasing Cr deposition, the 2DEG band becomes more clear and starts to show momentum splitting [Fig.~\ref{fig:doping}(e-j)] due to the formation of effective potential gradient across the sample surface. Such momentum-displaced Rashba effect derives from the spin-orbit coupling due to the effective potential gradient on the surface, $(\vec{k} \times \hat{z})\cdot \vec{\sigma}$, where $\vec{\sigma}$ is the Pauli matrices. The dispersion of the Rashba splitting bands can be approximated as
$E^\pm(k_\|)=E_0+{\hbar^2k_\|^2\over 2m^\ast}\pm\alpha |k_\||$,
where $m^\ast$ is the effective mass and $\alpha$ is the Rashba coupling parameter. Similar effect has been observed for various gas absorption or metal deposition on Bi$_2$Se$_3$ surface, e.g. CO \cite{HoffmanPRL11CO} and water \cite{WaterPRL2011}, nonmagnetic alkaline metals \cite{Valladoping}. From the evolution of electronic structure with Cr deposition, we extract the energy position of Dirac point E$_D$ [Fig. \ref{fig:doping}(k)] and the momentum splitting value $\Delta k_F$ [Fig.~\ref{fig:doping}(l)] after different deposition time.  Comparing the evolution of E$_D$ and $\Delta k_F$, we can see the Dirac point shifts downward quickly before the 2DEG band starts splitting after around 20 minutes of Cr deposition. This implies the deposited Cr atoms first donate electrons to Bi$_2$Se$_3$ sample and then form 2DEG with increasing effective electrical potential near the sample surface. The amount of charge transfer can be calculated from the change in the Fermi surface area including the SS, outer and inner Rashba bands using Luttinger's theorem \cite{Luttinger}. The charge transfer of Cr atoms  after 46 minutes Cr deposition is estimated to be $q \sim 0.06~ e^{-}$ per unit cell. The Rashba coupling parameter $\alpha$ can be extracted using $\alpha=\Delta k_F/2m^\ast$, where the $m^\ast$ is the effective mass of 2DEG. From Fig.~\ref{fig:doping}(l), the maximum coupling parameter is 0.34 eV$\cdot\mathrm{\AA}$, which is comparable with those of Au(111) surface \cite{LashellAu111, HofmannAu111} and 2DEG in semiconductors \cite{NittaInGaAs, DirkInAs}.

\begin{figure}
\includegraphics[width=1.0\linewidth] {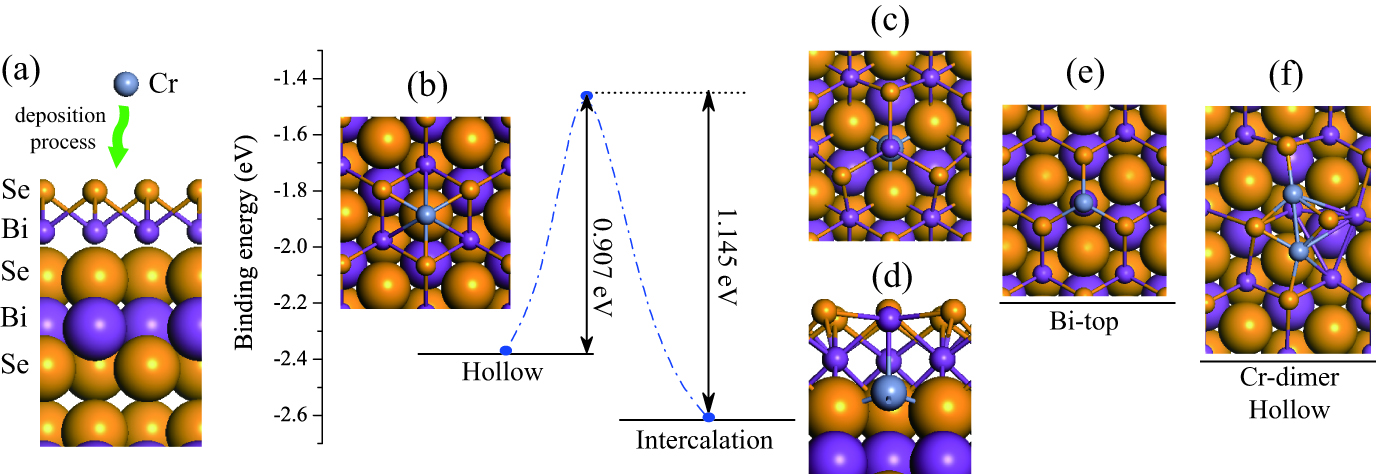}
\caption{(Color online) (a) A schematic diagram showing the deposition process of Cr atom to the surface of Bi$_2$Se$_3$. (b-f) Binding energies of Cr absorption on different sites: (b) hollow; (c,d) intercalation; (e) Bi-top; (f) Cr-dimer absorption on the hollow site. (b,c,e,f) are the top views and (d) is the side view. Black solid lines show the binding energies of absorption on different sites. The blue dash dot line stands for the chemical reaction path from climbing NEB calculation. The energy barrier is 0.907 eV.}
\label{fig:Binding}
\end{figure}

The absence of gap opening at the Dirac point and the emergence of Rashba-splitting bands suggest that the Cr atoms deposited on the surface interact with TI very differently from bulk doping \cite{XuePRL2014}. In order to understand why surface deposition of Cr atoms does not lead to the gap opening as in the case of bulk Cr doping, we use DFT to simulate the deposition process of Cr atoms on Bi$_2$Se$_3$(111) surface. At the beginning of the simulation, we put Cr atoms on different initial positions (including intercalation site, hollow site, bridge site, Bi-top site and Se-top site), and let them fully relax. It is found that Cr atoms on the Se-top site and bridge site will spontaneously relax to the Bi-top site. Therefore, we only have three energetically stable configurations as shown in Fig.~\ref{fig:Binding}. They are Cr absorptions on the hollow site [Fig. \ref{fig:Binding}(b)], intercalation site [Fig. \ref{fig:Binding}(c,d)], and Bi-top site [Fig.~\ref{fig:Binding}(e)], with binding energies ($E_{\rm b}$) \cite{note-Eb} of -2.378 eV, -2.631 eV and -2.200 eV respectively. Because $E_{\rm b}$ of all optimal structures are negative, the energy will be released during the deposition process, and the structures with absorbed Cr atoms are stable. Although the configuration for Cr atom on the intercalation site has the lowest $E_{\rm b}$, this structure is practically difficult to be realized. According to the climbing nudged elastic band (NEB) calculation \cite{henkelman2000climbing,*henkelman2000improved}, a large energy barrier (about 1 eV) should be overcome when Cr atom intercalates into the atomic layers underneath. Therefore, the most feasible configuration observed experimentally is Cr absorption on the hollow site with the relatively lower $E_{\rm b}$. This is very different from the substitutional Bi site for bulk Cr doping \cite{XuePRL2014}.

\begin{figure}
\includegraphics[width=0.7\linewidth] {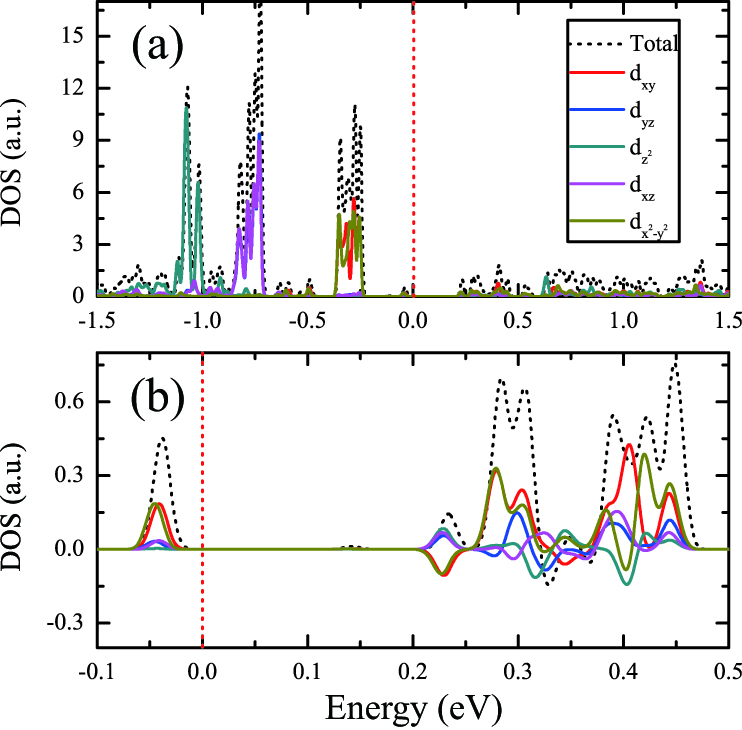}
\caption{(Color online) (a) The total and orbital-projected density of states (DOS) of Cr atom for the single adsorption on the hollow site. (b) The ${\sigma}_x$ component of DOS. The positive and negative stand for the spin-majority and spin-minority, respectively. The Fermi level is set to zero and marked by the red dot line.}
\label{fig:DOS}
\end{figure}

Here, we focus on the electronic structure of Cr-single absorption on the hollow site, and its density of state (DOS) is shown in Fig. \ref{fig:DOS}. Due to $C_{3v}$ symmetry around the absorption site, the crystal field splits Cr \textit{3d} orbitals into $A_{1}$ ($d_{z^2}$), $E_1$ ($d_{xz}$, $d_{yz}$) and $E_2$ ($d_{x^2-y^2}$, $d_{xy}$) states. Considering the Hunt's rule, a large energy splitting between the spin-majority states and the spin-minority states can be observed from our DFT calculations, and the five spin-majority states of Cr ($A_1^{\uparrow}$, $E_1^{\uparrow}$ and $E_2^{\uparrow}$) are fully occupied [see Fig. \ref{fig:DOS}(a)]. Due to the similar in-plane orientation, a strong \textit{p-d} hybridization between Se \textit{p} orbitals and Cr $d_{x^2-y^2}$ and $d_{xy}$ orbitals drives $E_2^{\uparrow}$ states close to the Fermi level, and the spin-polarized electron density becomes more extended which also could be found on the neighboring Se atoms. Therefore, the total magnetic moment of this structure is 5 $\mu_{\rm B}$.

For magnetic properties of Cr absorption on the hollow site, we calculate the magnetic anisotropy energy (MAE) from DFT simulations. It is found the magnetic easy axis is along the in-plane direction with MAE of 2.04 meV/atom. Based on the perturbation theory \cite{Bruno-PT-MAE,Daalderop-MAE}, around the Fermi level, MAE is mainly determined by the difference of SOC matrix elements ($H^{SO}_{ij}(\hat{n})\equiv \langle i|H^{SO}(\hat{n})|j \rangle$) along in-plane and out-of-plane direction, which can be expressed as $\Sigma_{i,j}\frac{1}{\Delta_{ij}}(|H^{SO}_{ij}(\hat{x})|^2-|H^{SO}_{ij}(\hat{z})|^2)$, where $i$ and $j$ stand for the occupied and un-occupied states respectively and $\Delta_{ij}$ is their energy difference. Shown in Fig. \ref{fig:DOS}(b), around the Fermi level, the states of Cr atom are mainly contributed by hybridized \textit{d} orbitals with opposite spins, and the coupling between them results in an in-plane magnetic easy axis \cite{Daalderop-MAE,note-MAE}. In TIs, the SS is protected by TRS and can only be gapped by out-of-plane ferromagnetism \cite{SCZhangQAH} or in-plane ferromagnetism combined with mirror symmetry breaking \cite{LiuCX-In-plane}. The absence of gap opening at the Dirac point in the ARPES data is in agreement with the in-plane magnetic easy axis \cite{BiSe-Fe-doping-2}.

If absorbed Cr atoms form in-plane ferromagnetism without breaking other symmetry, the effective exchange field will shift the Dirac point of SS without opening a band gap. The shift magnitude ($k_\|^{D}$) can be estimated by $k_\|^{D}\equiv m_{\hat{x}}/v_{F}$, in which $m_{\hat{x}}$ is the in-plane ferromagnetic coupling strength and $v_{F}$ is the Fermi velocity of Bi$_2$Se$_3$ ($v_{F}=3.55~\rm eV\cdot{\rm \AA}$ \cite{Kuroda-vf}). To estimate the shift of Dirac point, in our view, the maximum magnetic coupling strength should have the same order of magnitude with that of Cr-dimer absorbed on Bi$_2$Se$_3$(111) surface. Considering Cr-dimer absorption on the hollow site, calculated results are shown in Fig. \ref{fig:Binding}(f): the distance between two absorbed Cr atoms is 2.87 ${\rm \AA}$ and $E_{\rm b}$ is -2.409 eV. This value is slightly lower than that of single Cr absorption on hollow site [see Fig. \ref{fig:Binding}(b)], suggesting both structures can be observed in the experiment. The magnetic easy axis of Cr dimer is also along the in-plane direction with MAE of 3.32 meV/atom, and two Cr atoms prefer ferromagnetic coupling with coupling strength of 69.21 meV, and the average magnetic moment of each Cr atom is about 3.2 $\mu_{\rm B}$. If the in-plane ferromagnetic coupling formed in Cr deposited Bi$_2$Se$_3$ surface, using $k_\|^{D}\equiv m_{\hat{x}}/v_{F}$, we estimate the maximum momentum shift $k_\|^{D}$ is about 0.02 ${\rm \AA}^{-1}$, which is within our experimental uncertainty. We note that the absence of long range ferromagnetic order was reported \cite{BiSe-Fe-doping-2} in previous studies of surface deposition of Fe on  Bi$_2$Se$_3$ surface. Here, we tend to believe that there is no long-range ferromagnetic order down to 15 K for surface deposition of Cr, which needs to be confirmed by direct magnetic measurements.

In conclusion, our combined experimental and theoretical study show that, different from bulk Cr doped Bi$_2$Se$_3$, surface deposition of Cr atoms on Bi$_2$Se$_3$ shows no gap opening at Dirac point, implying that no out-of-plane ferromagnetism is formed on surface down to our measurement temperature of  15 K. According to DFT calculations, the difference of electronic and magnetic properties originates from the different Cr configurations: surface deposition on the hollow site versus substitutional site bulk doping. Due to the coupling between Cr \textit{3d} orbitals around the Fermi level, the magnetic easy axis of deposited Cr atoms is along the in-plane direction. In addition, deposited Cr atoms induce electron doping to the sample and form Rashba-splitting 2DEG bands near the sample surface. These results indicate the interaction between magnetic dopants and TIs strongly depends on the intricate configuration of magnetic atoms, which has important implication for manipulation of SS in TIs.

\begin{acknowledgement}

 This work is supported by the National Natural Science Foundation of China (grant No.~11274191 and 11334006) and Ministry of Education of China (20121087903, 20121778394).  E.Y.W. acknowledges support from the Advanced Light Source doctoral fellowship program. The Advanced Light Source is supported by the Director, Office of Science, Office of Basic Energy Sciences, of the U.S. Department of Energy under Contract No. DE-AC02-05CH11231. The crystal growth work at Purdue was supported by the DARPA MESO program (Grant No. N66001-11-1-4107)

\end{acknowledgement}

\section{Methods}
ARPES measurements were carried out at Beamline 12.0.1 of the Advanced Light Source with total energy resolution better than 25 meV. High quality bulk single crystal Bi$_2$Se$_3$ samples were cleaved {\it in situ} in UHV chamber with base pressure of 2.5$\times10^{-11}$ Torr before the measurements. The sample was kept at 15 K during Cr deposition and ARPES measurements. Cr was evaporated from a resistively heated tungsten basket, and the flux was calibrated by a quartz microbalance to be $\sim 0.17~ \AA$ per minute prior to experiment. Due to the different sticking coefficients between quartz and Bi$_2$Se$_3$(111) surface, the actual amount of Cr atoms absorbed on Bi$_2$Se$_3$ surface cannot be obtained directly.  In order to estimate the amount of Cr absorbed on Bi$_2$Se$_3$ surface, we compare the charge transfer from the Bader charge analysis \cite{henkelman2006-Bader} in the DFT calculation with the experimental charge transfer. The Bader charge analyis shows each Cr atom transfers 0.361 electron to Bi$_2$Se$_3$ surface. Experimentlly, using Luttinger's theorem \cite{Luttinger}, the total charge transfer from Cr atoms calculated from the change of Fermi surface area after deposition for 46 minutes is 0.06 electron per unit cell. Thus we estimate  $0.06/0.361 \sim 1/6$  Cr atom was absorbed per unit cell after 46 minutes deposition, and the corresponding coverage is $\sim 9\%$ ML Cr.

DFT calculations were carried out by using the projector augmented wave method \cite{PAW1994,PAW1999} and the generalized gradient approximation with Perdew-Burke-Ernzerhof type functional \cite{Perdew1996}, as implemented in the Vienna \textit{ab initio} simulation package \cite{Kresse1996}. Plane wave basis set with a kinetic energy cutoff of 340 eV was used. For the geometrical and electronic structures, a 3$\times$3$\times$1 supercell slab model with the in-plane experimental lattice constants was constructed, which contained 20 atomic layers (4 QL) of Bi$_2$Se$_3$, absorbed Cr atoms, and a vacuum layer larger than $20~\mathrm{\AA}$ along the \textit{z} direction. During the structure optimization, 16 atomic layers of Bi$_2$Se$_3$ on the bottom were fixed and the other atoms were fully relaxed until the residual force less than 0.01 eV/\AA. The Monkhorst-Pack \textit{k} points were 2$\times$2$\times$1. SOC was included to calculate the electronic structure \cite{Hobbs-SOC}. For the climbing nudged elastic band (NEB) calculations \cite{henkelman2000climbing,*henkelman2000improved}, the 3$\times$3$\times$1 supercell slab model contained 5 atomic layers (1 QL) of Bi$_2$Se$_3$, absorbed Cr atoms and a vacuum layer with fully relaxation.

\bibliography{Ref}

\begin{mcitethebibliography}{46}
\bibitem[not()]{note-Eb}
$E_{\rm b}$ is defined as $(E_{\rm t}-E_{\rm s}-n\mu_{\rm Cr})/n$, where
  $E_{\rm t}$ is the total energy of whole system, $E_{\rm s}$ and $\mu_{\rm
  Cr}$ are the energy of intrinsic Bi$_2$Se$_3$ thin film and the chemical
  potential of Cr atom respectively, and n is the number of Cr atoms.\relax
\mciteBstWouldAddEndPunctfalse
\mciteSetBstMidEndSepPunct{\mcitedefaultmidpunct}
{}{\mcitedefaultseppunct}\relax
\EndOfBibitem
\bibitem[not()]{note-MAE}
Here, below the Fermi level, the states of Cr are mainly contributed by
  spin-majority $E_2^{\uparrow}$ states; and above the Fermi level, the states
  of Cr are mainly contributed by spin-minority $E_2^{\downarrow}$ states and
  spin-majority $E_1^{\uparrow}$, $A_1^{\uparrow}$ states. The coupling of
  $E_2^{\uparrow}$ with $E_2^{\downarrow}$ and $E_1^{\uparrow}$ results in an
  in-plane magnetic easy axis; and the coupling of $E_2^{\uparrow}$ with
  $A_1^{\uparrow}$ is zero.\relax
\mciteBstWouldAddEndPunctfalse
\mciteSetBstMidEndSepPunct{\mcitedefaultmidpunct}
{}{\mcitedefaultseppunct}\relax
\EndOfBibitem
\end{mcitethebibliography}

\end{document}